\pdfoutput=1

\documentclass[11pt]{article}

\usepackage[final]{acl}

\usepackage{times}
\usepackage{latexsym}

\usepackage[T1]{fontenc}

\usepackage[utf8]{inputenc}

\usepackage{microtype}

\usepackage{inconsolata}

\usepackage{graphicx}
\usepackage{arydshln}
\usepackage{pifont}

\usepackage{booktabs}
\usepackage{tikz}
\usepackage{pgfplots}
\usepackage{multirow}
\usetikzlibrary{positioning,calc}%
\usetikzlibrary{backgrounds}
\usepackage{color, colortbl}
\usetikzlibrary{decorations.fractals}
\usetikzlibrary{patterns}
\usepackage{booktabs}
\usepackage[dvipsnames]{xcolor}
\usepackage{amsmath}
\usepackage{url}
\usepackage{amssymb}
\usepackage{hyperref}
\usepackage{makecell}
\usepackage[table]{xcolor}

\definecolor{myYellow}{rgb}{0.9,0.9,1}
\definecolor{WindowsBlue}{RGB}{2,152,219}
\definecolor{battleshipgrey}{rgb}{0.3, 0.3, 0.3}
\definecolor{brilliantrose}{rgb}{1.0, 0.33, 0.64}
\definecolor{americanrose}{rgb}{1.0, 0.01, 0.24}
\definecolor{jweigreen}{rgb}{0,0.45,0.24}
\definecolor{bluegray}{rgb}{0.1, 0.1, 0.4}
\definecolor{ao(english)}{rgb}{0.0, 0.5, 0.0}
\definecolor{blanchedalmond}{rgb}{1.0, 0.92, 0.8}
\definecolor{atomictangerine}{rgb}{1.0, 0.6, 0.4}
\definecolor{chocolate(web)}{rgb}{0.82, 0.41, 0.12}
\definecolor{bananayellow}{rgb}{1.0, 0.88, 0.21}
\definecolor{goldenbrown}{rgb}{0.6, 0.4, 0.08}
\definecolor{aliceblue}{rgb}{0.94, 0.97, 1.0}
\definecolor{beige}{rgb}{0.96, 0.96, 0.86}
\definecolor{babyblue}{rgb}{0.54, 0.81, 0.94}
\definecolor{camel}{rgb}{0.76, 0.6, 0.42}
\definecolor{cinnamon}{rgb}{0.82, 0.41, 0.12}

\title{Enhancing Speech Large Language Models with \\Prompt-Aware Mixture of Audio Encoders}

\author{Weiqiao Shan\textsuperscript{1}, Yuang Li\textsuperscript{2}, Yuhao Zhang\textsuperscript{3}, yingfeng luo\textsuperscript{1}, Chen Xu\textsuperscript{4}, Xiaofeng Zhao\textsuperscript{2}, \\ \textbf{Long Meng\textsuperscript{1}, Yunfei Lu\textsuperscript{2}, Min Zhang\textsuperscript{2}, Hao Yang\textsuperscript{2}, Tong Xiao\textsuperscript{1,5}\thanks{Corresponding author}, Jingbo Zhu\textsuperscript{1,5}}\\
\textsuperscript{1}School of Computer Science and Engineering, Northeastern University, Shenyang, China\\
\textsuperscript{2}Huawei Translation Services Center, Beijing, China\\
\textsuperscript{3}The Chinese University of Hong Kong, Shenzhen, China\\
\textsuperscript{4}College of Computer Science and Technology, Harbin Engineering University, Harbin, China\\
\textsuperscript{5}NiuTrans Research, Shenyang, China}

\begin{document}

\maketitle
\begin{abstract}
Connecting audio encoders with large language models (LLMs) allows the LLM to perform various audio understanding tasks, such as automatic speech recognition (ASR) and audio captioning (AC). Most research focuses on training an adapter layer to generate a unified audio feature for the LLM. However, different tasks may require distinct features that emphasize either semantic or acoustic aspects, making task-specific audio features more desirable. In this paper, we propose Prompt-aware Mixture (PaM) to enhance the Speech LLM that uses multiple audio encoders. Our approach involves using different experts to extract different features based on the prompt that indicates different tasks. Experiments demonstrate that with PaM, only one Speech LLM surpasses the best performances achieved by all single-encoder Speech LLMs on ASR, Speaker Number Verification, and AC tasks. PaM also outperforms other feature fusion baselines, such as concatenation and averaging. Our code
will be available at:~\url{https://github.com/shanweiqiao/PaM}
\end{abstract}
\section{Introduction}

Large language models (LLMs) have demonstrated exceptional performance across various natural language processing tasks~\cite{gpt2023gpt}, paving the way for developing multimodal models~\cite{li2023blip, xu2025qwen2, wang2024internvideo2}. In recent work, there has been a growing focus on merging speech encoders with LLMs, so that the LLM can understand the spoken content without the need for explicit transcription, promoting tasks such as direct speech translation~\cite{chen2023salm} and named entity recognition from speech~\cite{li2024usinglargelanguagemodel}. Much of this work leverages adapter layers like attention layers~\cite{yu2023connecting}, adaptive CTC downsamplers~\cite{ling2023adapting}, and convolutional layers~\cite{fathullah2023prompting} to downsample and map speech features into the LLM’s embedding space. Beyond semantic understanding tasks, Speech LLMs have been extended to encompass a broader range of applications, including audio event detection and audio captioning~\cite{Qwen2-Audio}.

\begin{figure}[tp]
  \centering
  \centerline{\includegraphics[scale=0.96]{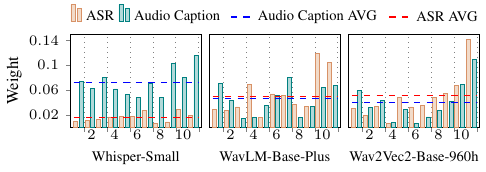}}
  \caption{ASR and Audio Caption tasks favor different encoders and layers of features. The x-axis corresponds to each layer of the encoder, while the bar chart illustrates the fine-grained layer importance, based on the normalized weight across layers from all encoders. The dotted lines indicate the average (AVG) importance of different encoders.}
  \vspace{-1.5em}
\label{fig:layer_importance}
\end{figure}

Multitasking requires that the input audio features contain as much relevant information as possible, representing the input speech, which may include speech content, noise, and speaker-specific characteristics. When fine-tuning self-supervised speech encoders, researchers assign learnable weights to each layer and observe that different downstream tasks prioritize different levels of features~\cite{chen2021wavlm}. In our Speech LLM framework, a similar trend is evident, where different tasks prioritize different encoders and feature levels (Figure~\ref{fig:layer_importance}). These biases arise from the inherent differences in the tasks themselves. For instance, the automatic speech recognition (ASR) task focuses solely on the speech content, disregarding other factors such as speaker characteristics and background noise. In contrast, tasks like audio captioning (AC) may rely on these additional factors that ASR intentionally excludes.

Consequently, researchers have proposed using multiple encoders to extract more robust features. For instance, WavLLM~\cite{hu2024wavllmrobustadaptivespeech} employs both the WavLM~\cite{chen2021wavlm} and the Whisper~\cite{radford2022robust} encoder, while SALMONN~\cite{tang2024salmonn} integrates the Whisper encoder and the BEATs~\cite{chen2023beats}. However, these approaches consider all encoders equally important and merge the features from different encoders based on a simple concatenation method across all tasks. As demonstrated in our experiments (Table~\ref{tab:main}), such conventional approaches can enhance performance in some tasks (e.g., audio captioning) but degrade others (e.g., ASR). Moreover, MoWE~\cite{zhang2024moweaudiomultitaskaudiollmsmixture} employs a strong encoder and multiple weaker encoders via the Mixture of Experts (MoE) approach. However, MoWE only utilizes the input audio to control the routing mechanism, without incorporating task-specific information in prompts, which leads to suboptimal results (Table~\ref{tab:more-results-based-on-routing-strategy}).
 
In this paper, we introduce Prompt-aware Mixture (PaM), a novel method based on MoE for merging multiple encoders to enhance Speech LLMs. Our approach integrates a prompt-aware routing mechanism, emphasizes feature fusion, and considers the relative importance of each encoder for different tasks, aiming to improve all downstream performance. PaM employs three distinct audio encoders: the Whisper encoder, WavLM, and Wav2Vec2~\cite{baevski2020wav2vec}. We train a set of experts for prompt-aware feature fusion, comprising one shared expert and four task-specific experts. On each task, an expert learns the optimal weights for each encoder and its respective layers, and subsequently maps the resulting features to the embedding space of the Qwen2.5 model~\cite{qwen2.5}. The embedding of the prompt is utilized to determine the appropriate routing. Notably, in PaM, the routing guides the selection of the fusion parameters rather than the choice of the encoder. Experiments are conducted across three tasks: ASR, speaker number verification (SNV), and AC. On all datasets, including LibriSpeech~\cite{panayotov2015librispeech}, AMI~\cite{kraaij2005ami}, AIR-Bench (SNV)~\cite{yang2024air}, and AudioCaps~\cite{audiocaps}, PaM achieves relative improvements of 15\%, 25\%, 3.4\%, and 7.6\%, respectively, in comparison to the best-performing single-encoder Speech LLM, and achieves a better average rank on all tasks compared with conventional approaches. Our contributions can be summarized as follows:
\begin{itemize}
\item We propose a novel multi-encoder Speech LLM, which effectively leverages features from every layer of each encoder.

\item We introduce PaM, a method based on MoE that incorporates a prompt-aware routing mechanism to assign distinct weights to each encoder and its layers based on the task.

\item We conducted comprehensive experiments demonstrating that PaM significantly enhances the overall performance of all downstream tasks. Additionally, we present detailed feature importance analyses and explore various combinations of speech encoders and LLMs. 
\end{itemize}

\begin{figure*}[t]
\centering
  \centerline{\includegraphics[scale=1.04]{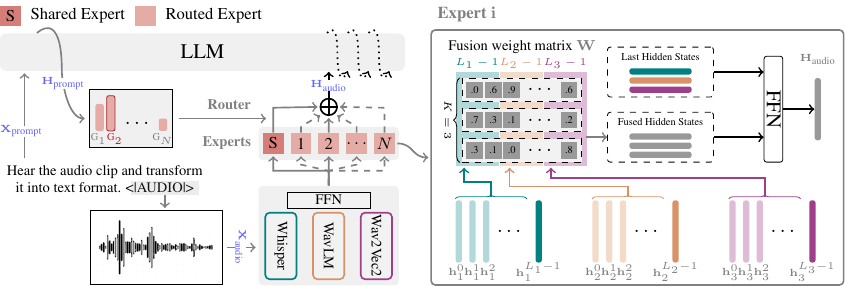}}
  \caption{The architecture of the proposed PaM method. The output feature $\mathbf{H}_\text{prompt}$ of the prompt $\mathbf{X}_\text{prompt}$ guides the routing of the MoE structure adapter, which incorporates a fixed shared expert (denoted as expert S) and a single routed expert (denoted as expert $i\in[1,N]$, where $N$ represents the total number of routed experts). For each expert, the last hidden states from all encoders $\mathbf{h}_i^{L_i - 1}$ are concatenated with $K$ fused hidden states ($K$ being a hyperparameter, with the default value $K=3$) derived from a fusion weight matrix $\mathbf{W}$. Subsequently, a feedforward network (FFN) is applied to align with the dimensions of the LLM.}
  \label{fig:moe-model-arch}
\end{figure*}

\section{Method}

In this section, we begin with an overview of the proposed PaM method (Figure~\ref{fig:moe-model-arch}). We then elaborate on the details of the encoder fusion process executed by a single expert, and describe the prompt-aware routing method.

\subsection{Overall Architecture}

The architecture of the proposed PaM method is depicted in Figure~\ref{fig:moe-model-arch} (left). As described in Equation~\ref{eq:1}, the LLM accepts the text prompt $\mathbf{X}_\text{prompt}$, which includes task-related information, along with the speech features $\mathbf{H}_\text{audio}$ as input, and subsequently generates the response $\mathbf{Y}$.
 \begin{align}
 \label{eq:1}
 \mathbf{Y} = \text{LLM}(\mathbf{X}_\text{prompt}, \mathbf{H}_\text{audio})
\end{align}

\noindent To obtain $\mathbf{H}_\text{audio}$, we employ three encoders: the Whisper encoder, WavLM, and Wav2Vec2. For each encoder, the hidden states are initially processed by a feed-forward network (FFN) as described in Equation~\ref{eq:2}\footnote{The FFN module projects the hidden states from the encoder's dimension to the LLM's dimension, and maps the features from each encoder into a unified space that is shared across all encoders.}, resulting in the feature of each encoder, denoted as $\mathbf{H}_{i}$.
 \begin{align}
 \label{eq:2}
 \mathbf{H}_i = \text{FFN}_i(\text{Encoder}_i(\mathbf{X}_\text{audio}))
\end{align}

\noindent Next, as shown in Equation~\ref{eq:3}, we combine these features using a fusion method based on MoE, which includes a shared expert and \textit{N} routed experts, where the number of routed experts corresponds to the number of predefined tasks. During inference, only one routed expert is selected, based on the task indicated by the prompt. 
 \begin{align}
 \label{eq:3}
  &\mathbf{H}_\text{audio} = \text{Expert}_\text{share}(\mathbf{H}_{\{1,2,3\}}) \nonumber \\&+ \sum_{j=1}^{N}\text{G}_j(\mathbf{X}_\text{prompt})\times\text{Expert}_j(\mathbf{H}_{\{1,2,3\}})
\end{align}

\noindent Overall, each expert processes the features from all encoders ($\mathbf{H}_{\{1,2,3\}}$) for feature fusion. The shared expert extracts common features for all tasks, while the routed expert performs task-specific feature fusion. The routing is determined by the user input, which is the prompt.

\subsection{Multi-layer Fusion}

We describe the multi-layer fusion process in Figure~\ref{fig:moe-model-arch} (right). Different encoders exhibit distinct strengths. For example, WavLM is excellent at extracting speaker information~\cite{chen2021wavlm}, while Wav2Vec2 excels in capturing semantic content~\cite{baevski2020wav2vec}. The Whisper encoder~\cite{radford2022robust}, trained on a vast amount of data, provides superior features for AC and ASR in noisy environments\footnote{These biases in the ability of different encoders on different tasks are also consistent with the results of our single-encoder baselines shown in Table~\ref{tab:main}}. Additionally, features from different layers contain varying levels of information. Deeper layers hold high-level semantic information, whereas lower layers may contain fine-grained acoustic details. Thus, for feature fusion, we consider features from all layers of all three encoders. Specifically, for the feature $\mathbf{H}_i$ from a single encoder, it includes hidden states from all $L_i$ Transformer~\cite{NIPS2017_3f5ee243} layers as well as $\mathbf{h}_i^{0}$, the hidden states following the convolutional layers (Equation~\ref{eq:4}).
\begin{align}
 \label{eq:4}
 \mathbf{H}_{i} = \{\mathbf{h}_i^{0}, \mathbf{h}_i^{1}, \ldots \mathbf{h}_i^{L_i - 1}\}
\end{align}

\noindent As illustrated in Equation~\ref{eq:5}, we consider the hidden states $\mathbf{h}_i^{\{0,1,\ldots, (L_i\text{-}2)\}}$ to derive the fused hidden states $\mathbf{h}_{k}^\text{fused}$. For each hidden state $\mathbf{h}^l_{i}$ of all three encoders, a set of scalar weights containing $\sum_i^3 (L_i - 1)$ elements is assigned to control its relative importance. We denote a set of scalar weights as $\{w_{i,k}^{l} | i \in [1,3], l \in [1, L_i - 1]\}$ and then use it to generate fused hidden states $\mathbf{h}_{k}^\text{fused}$. To maintain the diversity of the fusion feature, we utilize $K$ sets of scalar weights. Thus, the dimension of the whole learnable matrices is $\mathbf{W} \in \mathbb{R}^{K\times(\sum_i^3 (L_i - 1))}$.
\begin{align}
 \label{eq:5}
 \mathbf{h}_{k}^\text{fused} = \sum_{i=1}^3 \sum_{l=1}^{L_i - 1} w_{i,k}^l \cdot \mathbf{h}^l_{i}
\end{align}

\noindent Finally, we concatenate the last hidden states of the three encoders $\mathcal{H}^\text{last} = \{\mathbf{h}_{i}^{L_i - 1} | i \in [1,3]\}$ with the $K$ fused hidden states $\mathbf{h}_{\{1,\ldots,K\}}^\text{fused}$ along the feature dimension. Afterward, we apply an FFN to compress the feature dimension to match the dimension of the LLM embedding (Equation~\ref{eq:6})\footnote{We leverage both the final and fused hidden states, which are commonly used in semantic-related tasks (e.g., ASR) and acoustic-related tasks (e.g., audio captioning). The subsequent FFN module projects the concatenated features to the dimension of the LLM input embeddings, ensuring alignment between the feature space of speech features and text features.}.
\begin{align}
 \label{eq:6}
 \mathbf{h}^{\text{final}} &= \text{Concat}(\mathcal{H}^\text{last}, \mathbf{h}_{\{1,\ldots,K\}}^\text{fused}) \nonumber \\ 
 \text{Expert}(\cdot) &= \text{FFN}(\mathbf{h}^{\text{final}})
\end{align}

\noindent The parameters in our fusion method are independent among the routed experts. The use of fusion weight matrices highlights multi-level feature fusion, while the final concatenation followed by the FFN provides more fine-grained feature fusion.

\subsection{Prompt Aware Routing}

A prompt refers to a text segment that provides context or objectives for generation, which can typically be categorized into several distinct types according to task. For instance, speech-related tasks, sound-related tasks, and speech chat tasks~\cite{yang2024air}. In this paper, we investigate three tasks: ASR, SNV, and AC. We utilize distinct experts for each task. For effective routing, the router must identify the task type based on the prompt. We employ a simple classification approach (Equation~\ref{eq:7} and~\ref{eq:8}) wherein we use the last hidden states $\mathbf{H}_\text{prompt}$ of the prompt from the LLM, followed by an FFN and Softmax activation, to obtain the task posteriors $\text{P}(\text{Task}|\mathbf{H}_\text{prompt})$.
\begin{align}
 \label{eq:7}
 \mathbf{H}_\text{prompt} &= \text{LLM}(\mathbf{X}_\text{prompt})\\ 
 \label{eq:8}
 \text{P}(\text{Task}|\mathbf{H}_\text{prompt}) &= \text{Softmax}(\text{FFN} (\mathbf{H}_\text{prompt}))
\end{align}

\noindent As shown in Equation~\ref{eq:9}, we select the routed expert with the Top-1 probability by the indicator function.
\begin{align}
 \label{eq:9}
 \text{G}_j = 
 \begin{cases}
 1 & \text{if } j\in\text{Top-1}(\text{P}(\text{Task}|\mathbf{H}_\text{prompt}))\\
 0 & \text{otherwise}
 \end{cases}
\end{align}

To train the FFN, we create diverse prompts for each task using the LLM. Specifically, we manually write several prompts for each task and instruct ChatGPT to rewrite these prompts. We list the examples of these prompts in Appendix~\ref{sec:appendix-prompt-example}. It is important to note that the audio features follow the prompt because we use the prompt to guide feature extraction and fusion. This approach differs from other works, where the audio features <|AUDIO|> can be positioned before the prompt.

\subsection{Training Objective}

The training loss function, as illustrated in Equation~\ref{eq:10}, is the sum of the cross-entropy loss $\mathcal{L}_\text{G}$ for prompt-aware routing and the cross-entropy loss $\mathcal{L}_\text{llm}$ between the LLM's output $\mathbf{Y}$ and the ground truth $\hat{\mathbf{Y}}$.
$\hat{\mathbf{Y}}$.
 \begin{align}
 \label{eq:10}
 \mathcal{L} &= \mathcal{L}_\text{G}(\text{P}(\text{Task}|\mathbf{H}_\text{prompt}), \text{Task}) \nonumber \\ 
 &+ \mathcal{L}_\text{llm}(\mathbf{Y}, \hat{\mathbf{Y}}) 
\end{align}

\section{Experimental Setups}

\subsection{Datasets and Evaluation Metrics}

We assess the efficacy of our method across three audio-to-text tasks: automatic speech recognition (ASR), speaker number verification (SNV), and audio captioning (AC). We list the detailed information on training data in Appendix~\ref{sec:appendix-datas-and-model}. In total, the training data contains 450 hours of audio signals. The test dataset includes LibriSpeech-test-clean, LibriSpeech-test-other, AMI, the SNV test set from AIR-Bench, and the test set of AudioCaps along with its corresponding QA version from AudioBench~\cite{wang2024audiobench}, which contains diverse questions. ASR tasks focus on semantics. The LibriSpeech test set originates from audiobooks, demonstrating ASR performance in a clean scenario. AMI, a real meeting corpus containing spontaneous talk, reflects ASR performance in a more challenging, real-world scenario. SNV and AC test sets can indicate the Speech LLM's ability to understand speaker and acoustic information. We evaluate the performance using word error rate (WER) for ASR tasks, accuracy for SNV, and METEOR~\cite{banerjee2005meteor} for AC. Additionally, we list the results on AC and AC QA tasks with more metrics in Appendix~\ref{sec:appendix-AC-with-more-metrics}.

\begin{table*}[htp]
\centering
\resizebox{\textwidth}{!}{\begin{tabular}{l l l l l l l c}
\toprule
\multirow{2}{*}{Model} & \multicolumn{2}{c}{LibriSpeech} & AMI & SNV & AudioCaps & AudioCaps QA & \multirow{2}{*}{AVG Rank$\downarrow$} \\ 
\cmidrule(r){2-7}
 & WER(clean)$\downarrow$ & WER(other)$\downarrow$ & WER$\downarrow$ & Acc$\uparrow$ & METEOR$\uparrow$ & METEOR$\uparrow$ & \\ 
\midrule
\rowcolor{black!6} Single-encoder Baselines & & & & & & & \\
\quad- Whisper~\cite{radford2022robust} & 9.61 & 16.73 & \textbf{16.27} & 18.8\% & \textbf{32.96} & \textbf{15.04} & 5.67 \\
\quad- WavLM~\cite{chen2021wavlm} & 5.59 & 10.57 & 18.97 & \textbf{41.4\%} & 27.14 & 12.77 & 5.50 \\
\quad- Wav2Vec2~\cite{baevski2020wav2vec} & \textbf{4.30} & \phantom{0}\textbf{9.46} & 26.69 & 39.0\% & 23.81 & 11.20 & 5.33 \\
\midrule
\rowcolor{black!6} Multi-encoder Baselines & & & & & & & \\
\quad- WavLLM~\cite{hu2024wavllmrobustadaptivespeech} & 4.95 (\textcolor{red!70!black}{-\;0.65}) & \phantom{0}9.19 (\textcolor{green!70!black}{+0.27}) & 15.29 (\textcolor{green!70!black}{+0.98}) & 39.2\% (\textcolor{red!70!black}{-\;2.20}) & 34.93 (\textcolor{green!70!black}{+1.97}) & 16.35 (\textcolor{green!70!black}{+1.31}) & 2.67 \\
\quad- SALMONN~\cite{tang2024salmonn} & 5.04 (\textcolor{red!70!black}{-\;0.74}) & \phantom{0}9.70 (\textcolor{red!70!black}{-\;0.24}) & 19.04 (\textcolor{red!70!black}{-\;2.77}) & 49.4\% (\textcolor{green!70!black}{+8.00}) & 34.86 (\textcolor{green!70!black}{+1.90}) & 15.97 (\textcolor{green!70!black}{+0.93}) & 3.50 \\
\quad- Average & 4.76 (\textcolor{red!70!black}{-\;0.46}) & 10.43 (\textcolor{red!70!black}{-\;0.97}) & 17.20 (\textcolor{red!70!black}{-\;0.93}) & 45.5\% (\textcolor{green!70!black}{+4.10}) & 33.22 (\textcolor{green!70!black}{+0.26}) & 15.53 (\textcolor{green!70!black}{+0.49}) & 3.67 \\
\midrule
\rowcolor{black!6} PaM (Ours) & 3.65 (\textcolor{green!70!black}{+0.65}) & \phantom{0}7.07 (\textcolor{green!70!black}{+2.39}) & 12.79 (\textcolor{green!70!black}{+3.48}) & 42.8\% (\textcolor{green!70!black}{+1.40}) & 35.47 (\textcolor{green!70!black}{+2.51}) & 15.70 (\textcolor{green!70!black}{+0.66}) & \textbf{1.67} \\
\bottomrule
\end{tabular}}
\caption{Comparison of the proposed PaM method with single and multi-encoder baselines. Values in the brackets indicate performance improvement (green) or degradation (red) compared to the best single encoder result. The AVG rank column shows the average rank on each task. Smaller ranks indicate better performance.}
\label{tab:main}
\end{table*}

\subsection{Model Architecture and Training}

We train our model based on Huggingface Transformers Library\footnote{\url{https://github.com/huggingface/transformers}}. Our model consists of three audio encoders, a pre-fusion adapter for each encoder, a PaM fusion module, and an LLM. In our main experiments, the encoders are Whisper-Small encoder, WavLM-Base-Plus, and Wav2Vec2-Base-960h\footnote{The links to the pretrained models and datasets used can be found in Appendix~\ref{sec:appendix-datas-and-model}. The implementation details of baseline methods are available in Appendix~\ref{sec:appendix-baseline-implement}.}, each with approximately 100 million parameters. We downsample the features from the Whisper encoder by a factor of two, resulting in a frame length of 40ms, consistent with the frame length of Wav2Vec2 and WavLM. The pre-fusion adapter is an FFN that transforms the encoder's hidden dimension $D_{\text{E}}$ to the LLM's hidden dimension $D_{\text{LLM}}$. Each expert in the PaM fusion module includes a fusion weight matrix ($\mathbb{R}^{3L\times3}$) and a linear layer ($\mathbb{R}^{6D_{\text{LLM}}\times D_{\text{LLM}}}$) to fuse features from all encoders. Here, $L$ represents the number of layers in the encoder, which is 12 for all encoders in our experiments. For the fused features, we set $K=3$, corresponding to the number of last hidden states. We utilize four routed experts, each corresponding to a specific task category: ASR-clean, ASR-noisy, SNV, and AC. For each category, we generate 50 prompts using ChatGPT~\cite{gpt2023gpt}\footnote{We provide a few examples of the prompts we used in Appendix~\ref{sec:appendix-prompts-in-training-inference}}. For the LLM model, we select the Qwen2.5-3B~\cite{qwen2.5}. In Section~\ref{sec:results}, we also experiment with other encoders, including Hubert-Base-LS960, Whisper-Large-v3, and WavLM-Large.\footnote{Additionally, we add the BEATs model, which performs well on the AC task, to further enhance PaM in Appendix~\ref{sec:appendix-pam-with-more-encoders}.}

We list the training and inference parameters in Appendix~\ref{sec:appendix-training-inference-param}.

\begin{table*}[t]
\scriptsize
\centering
\resizebox{\textwidth}{!}{\begin{tabular}{l c c c c c c c}
\toprule
\multirow{2}{*}{Encoders} & \multicolumn{2}{c}{LibriSpeech} & AMI & SNV & AudioCaps & AudioCaps QA & \multirow{2}{*}{AVG Rank$\downarrow$} \\ 
\cmidrule(r){2-7}
& WER(clean)$\downarrow$ & WER(other)$\downarrow$ & WER$\downarrow$ & Acc$\uparrow$ & METEOR$\uparrow$ & METEOR$\uparrow$ & \\ 
\midrule
 Whisper+X & 4.42 & 8.92 & 13.96 & 43.70\% & \textbf{35.41} & 16.11 & 4.5 \\
WavLM+X & 4.07 & 7.97 & 18.47 & 47.47\% & 32.48 & 15.01 & 5.5 \\
Wav2Vec2+X & \textbf{3.42} & 7.20 & 18.18 & 45.13\% & 32.33 & 15.02 & 4.3 \\
HuBERT+X & 3.87 & 8.21 & 19.49 & 36.90\% & 31.82 & 15.08 & 6.8 \\
\midrule
Whisper+X+Y & 4.22 & 8.11 & \textbf{13.79} & \textbf{49.77\%} & 35.37 & \textbf{16.31} & \textbf{3.0} \\
WavLM+X+Y & 3.72 & 7.99 & 16.35 & 38.73\% & 34.23 & 15.82 & 4.0 \\
Wav2Vec2+X+Y & 3.77 & \textbf{6.90} & 16.90 & 42.63\% & 34.23 & 15.82 & 3.8 \\
HuBERT+X+Y & 4.03 & 7.96 & 17.13 & 44.17\% & 34.18 & 16.13 & 4.0 \\
\bottomrule
\end{tabular}}
\caption{Results with different combinations of encoders. The first and last four rows represent combinations of two and three encoders respectively. Each row's results are the average performance of a fixed encoder paired with all possible combinations of one or two other encoders, highlighting the unique strengths of each encoder.}
\label{tab:multiencoder}
\end{table*}

\section{Results}\label{sec:results}

\subsection{Main Results}

As demonstrated in Table~\ref{tab:main}, we compare the proposed PaM method with single and multi-encoder baselines. Each encoder exhibits distinct advantages. When utilizing a single encoder, the Speech LLM with the Whisper encoder performs best on the AMI dataset and AC tasks. The primary reason is that the Whisper model is trained on vast speech data, exposing it to diverse acoustic conditions. Consequently, it excels in challenging ASR and AC tasks in real-world environments and noisy conditions. The WavLM encoder, trained on multi-speaker speech signals, provides the best features for the SNV task. The Wav2Vec2 encoder performs best on the LibriSpeech dataset mainly because it was pretrained on this dataset. However, since the LibriSpeech dataset consists of clean audiobooks, the Speech LLM with the Wav2Vec2 encoder shows poor performance on the AMI and AudioCap datasets.

We reimplemented the feature fusion methods of WavLLM and SALMONN, training the Speech LLM with the three audio encoders in our setups. Both methods use concatenation but are followed by different projection layers: WavLLM with a linear layer and SALMONN with a Q-former layer. Additionally, we implemented a simple averaging method that directly computes the average of the features from the three encoders. Compared to the best performance of single encoder baselines, all three fusion methods achieve better METEOR scores on AC tasks. However, performance may degrade on other tasks. For example, we observed performance degradation for all three methods on the LibriSpeech test-clean subset. This is expected since the same features are used for all tasks. Features containing more useful acoustic information for AC tasks may lack useful semantic information for ASR tasks.

PaM consistently outperforms all single encoder baselines, delivering performance improvements across all tasks. This consistent improvement can be attributed to the MoE structure adapter, which provides unique features tailored for each task. Compared to other fusion methods (i.e., concatenation and averaging), PaM achieves significantly lower WERs on the LibriSpeech and AMI datasets and similar performance on SNV and AC tasks.

\subsection{Feature Importance}

\begin{figure}[t]
\centering
  \centerline{\includegraphics[scale=1.0]{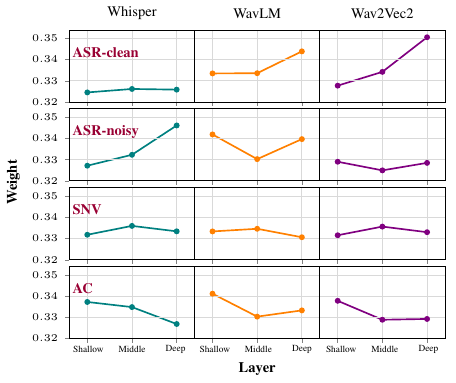}}
  \caption{The weights for each encoder and its layers.}
  \vspace{-1.5em}
  \label{fig:hidden-layer-importance}
\end{figure}

In Figure~\ref{fig:hidden-layer-importance}, we visualize the fusion weights for each expert, excluding the shared expert, which can be interpreted as the fusion weight for each task. We sum the weights for every four layers to enhance clarity, resulting in the total weight for shallow, middle, and deep layers. Generally, different tasks require different features, so each expert has distinct fusion weights. Specifically, when the expert for ASR-clean is activated, it mainly focuses on features from WavLM and Wav2Vec2, especially the deep layers. When the expert for ASR-noise is activated, it primarily focuses on features from the Whisper encoder and WavLM. For SNV and AC tasks, all three encoders have similar fusion weights. For the SNV task, features from the middle layers are more important, while for the AC task, shallow layers contribute more.

\subsection{Combinations of Encoders}

In Table~\ref{tab:multiencoder}, we extend our investigation to encompass more combinations of encoders, including two and three encoders, and incorporate the HuBERT encoder. To highlight the strengths of each encoder, we calculate the performance by keeping one encoder fixed and varying the other encoders, then computing the average. The average (AVG) rank reflects the overall performance across multitasks. It is evident that using three encoders significantly outperforms using two encoders in all combinations, thereby demonstrating the effectiveness of employing more encoders for Speech LLMs.
 
We can observe that different encoders offer varying benefits, which proves that the task-specific encoders can significantly improve the performance on the corresponding tasks. For example, when Whisper is used, regardless of how many encoders are employed, the Speech LLM achieves the lowest WER on AMI and the highest METEOR scores on AudioCaps. On the other hand, Wav2Vec2 provides an advantage for recognizing speech signals in LibriSpeech. This indicates that when selecting encoders for Speech LLM, it is essential to consider the domain, downstream tasks, and the capabilities of each encoder. We suggest using a robust general domain model like Whisper in combination with domain-specific encoders such as Wav2Vec2.

\begin{table*}[t]
\tiny
\centering
\resizebox{\textwidth}{!}{\begin{tabular}{l c c c c c c}
\toprule
\multirow{2}{*}{\makecell{Models}} & \multicolumn{2}{c}{LibriSpeech} & AMI & SNV & AudioCaps & AudioCaps QA \\ 
\cmidrule(r){2-7}
& WER(clean)$\downarrow$ & WER(other)$\downarrow$ & WER$\downarrow$ & Acc$\uparrow$ & METEOR$\uparrow$ & METEOR$\uparrow$ \\ 
\midrule
Base PaM & 3.65 & \phantom{0}7.07 & 12.79 & 42.8\% & 35.47 & 15.70 \\
\midrule
\rowcolor{black!6} PaM with Larger Encoders & & & & & & \\
\quad- \ding{172} Whisper-Medium & 3.93 & \phantom{0}7.93 & 12.43 & 47.5\% & 35.61 & 15.58 \\
\quad- \ding{173} WavLM-Large & 3.75 & \phantom{0}6.43 & 12.10 & 45.6\% & 35.95 & 16.71 \\
\quad- \ding{174} Wav2Vec2-Large & 3.74 & \phantom{0}6.49 & 15.06 & 47.6\% & 35.50 & \textbf{16.74} \\
\quad- \ding{172} + \ding{173} + \ding{174} & \textbf{3.58} & \phantom{0}\textbf{5.93} & \textbf{11.51} & \textbf{56.9\%} & \textbf{36.94} & 16.50 \\
\midrule
\rowcolor{black!6} PaM with Different LLMs & & & & & & \\
\quad- Qwen2.5-7B & 3.68 & \phantom{0}8.36 & 15.26 & 43.7\% & 35.46 & 15.71 \\
\quad- LLaMA3.2-3B & 4.98 & 11.57 & 15.57 & 50.5\% & 35.83 & 16.34 \\
\quad- LLaMA3.1-8B & 4.85 & \phantom{0}8.87 & 15.01 & 50.8\% & 35.81 & 15.82 \\
\bottomrule
\end{tabular}}
\caption{Results with larger encoders and various LLMs. To enhance performance, we replaced the Base version's encoders and experimented with different LLMs.}
\label{tab:llm}
\end{table*}

\subsection{Larger Encoders and LLMs}

We aim to further enhance performance by replacing the encoders in the proposed method with their larger versions (Table~\ref{tab:llm}). Specifically, we replace the Whisper-Small encoder with the Whisper-Medium encoder, Wav2Vec2-Base-960h with Wav2Vec2-Large-960h, and WavLM-Base-Plus with WavLM-Large. Our observations indicate that on the LibriSpeech-clean dataset, performance does not significantly improve and may even slightly degrade. However, for the SNV and AC tasks, performance consistently improves, suggesting that more challenging sound-related tasks benefit more from better encoders. Additionally, we observe that when all encoders are replaced with their larger versions, we achieve the best performance across almost all tasks, albeit at the cost of increased computation.

In our investigation of other LLMs, including Qwen2.5-7B, LLaMA3.2-3B, and LLaMA3.1-8B, we observed some improvements in certain tasks. However, the overall performance was not superior to that of Qwen2.5-3B. The potential reason for this is that we used short audios samples, and both the prompts and answers were brief, thereby not fully utilizing the strong semantic understanding capabilities of the larger LLMs. Consequently, we opted to use Qwen2.5-3B in this paper. It is important to emphasize that for Speech LLMs, the extracted features may be more critical than the LLM itself for many downstream tasks. In addition, we also compare the concatenation fusion strategy (WavLLM) and PaM using larger LLMs based on the LLaMA3.1-8B and Qwen2.5-7B. We find that the performance of the concatenation fusion strategy is inconsistent across different base models of similar size, whereas PaM maintains stability, as detailed in Appendix~\ref{sec:appendix-wavllm-and-pam-larger-llm}.

\subsection{Parameters of the Adapter}

In PaM, we employ multiple experts, merge and concatenate various features. Consequently, the number of parameters is slightly higher than that of the baseline concatenation and average methods. To ensure a fair comparison, we reduce the dimensionality within PaM, resulting in only 29M total parameters, similar to the baselines. PaM outperforms the baseline across almost all tasks, with similar overall parameters. Notably, during inference, PaM activates only 26M parameters, in contrast to the 37M parameters activated by the concatenation method, demonstrating the efficiency of PaM. In this configuration, each expert contains only 0.9M parameters, which is smaller than other components of the model, such as the LLM and encoders. Consequently, PaM can be further enhanced by increasing the number of experts with minimal impact on computational cost.

\begin{figure}[tp]
  \centering
  \centerline{\includegraphics[scale=1.0]{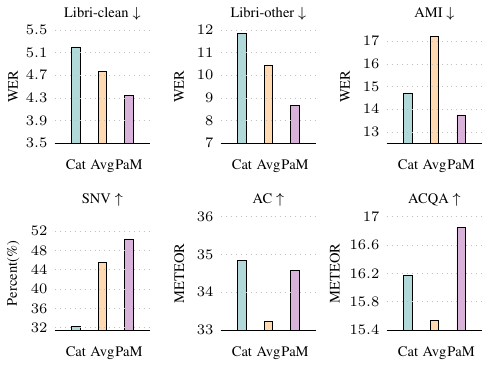}}
  \caption{Performance comparison of a smaller PaM (29M parameters) with Concatenation (37M parameters) and Average (24M parameters).}
\label{fig:small}
\end{figure}

\section{Discussions}

\textbf{Ablation study and routing method}: To further validate the effectiveness of PaM, we conducted an ablation study, such as PaM without the shared expert, as detailed in Appendix~\ref{sec:appendix-ablation-routing}. We also compare PaM against a learnable routing approach without task information, commonly employed in MoE models. The results indicate that PaM is more efficient and better suited to handling multiple downstream tasks than conventional routing and fusion strategies. Moreover, Table 3 highlights the strengths of each encoder. These findings encourage researchers to select encoders tailored to their downstream tasks for the Speech LLM. While we did not incorporate this prior knowledge into the routing mechanism, leveraging it presents a promising direction for future work.

\noindent \textbf{Other further work}: \textbf{1.More and Unseen Tasks}. Although our experiments involve three tasks and five datasets, they are representative as they encompass both semantic and acoustic-related tasks. We believe PaM can be extended to other tasks, which we will validate in future work. \textbf{2.Leverage Multimodal LLMs}. Features across various encoders hinder initialization from pretrained multimodal LLMs in our work (Appendix~\ref{sec:appendix-init-setup-for-LLM}). We will explore strategies to more effectively leverage pretrained multimodal LLMs, such as~\cite{lai2024lisa}. \textbf{3.Efficiency}. Improving the efficiency of LLMs has attracted much attention. Enhancing the computational efficiency of speech LLMs presents another promising avenue for exploration (Appendix~\ref{sec:appendix-init-setup-for-LLM}).

\section{Related Works}

\paragraph{Audio Encoders:} Audio encoders can be classified into supervised and self-supervised models. Supervised models typically employ ASR tasks to train an end-to-end model with an audio encoder and a text decoder. By omitting the decoder, the encoder can serve as a feature extractor~\cite{radford2022robust, baevski2020wav2vec}. Self-supervised models can be trained on unlabeled speech signals. For instance, Wav2Vec2~\cite{baevski2020wav2vec} and HuBERT~\cite{hsu2021hubert} were trained to predict the pseudo-discrete targets at masked time steps. WavLM~\cite{chen2021wavlm} is a variant of HuBERT, designed to facilitate speaker identity extraction by using multi-speaker signals. Different model architectures, training methods, and data can result in encoders with distinct properties and advantages, making the mixture of audio encoders effective for Speech LLMs.

\paragraph{Speech LLM:} To construct end-to-end speech LLMs, a natural approach is to extract discrete tokens from continuous speech signals and then expand the vocabulary of text LLMs to understand these speech tokens~\cite{rubenstein2023audiopalm, veluri2024beyond, ma2024language}. An alternative is to use an adapter layer to directly convert the continuous speech features into the continuous embedding space of the LLM. For example, QwenAudio~\cite{Qwen2-Audio} employs average pooling to downsample speech features, followed by two linear layers for projection. SALMONN~\cite{tang2024salmonn} utilizes the Q-former~\cite{yu2023connecting}, a cross-attention-based adapter, to achieve a higher compression ratio. In parallel, to achieve high compression, Soundwave replaces cross-attention with a lightweight self-attention module that treats the inputs as queries~\cite{zhang2025soundwave}. Compared to previous works, our adapter handles more encoders and generates different features based on the prompt, rather than a single feature for all prompts.

\paragraph{Mixture of Experts:} MoE has attracted growing interest, which replaces the FFN sub-layer in Transformer models with multiple experts~\cite{shazeer2017outrageously}. These MoE methods typically employ massive experts and extremely sparse activation routing, increasing model size while maintaining constant inference costs, without explicitly considering the specialization of individual experts~\cite{fedus2022switch, lepikhin2020gshard}. However, the vast scale of these models presents significant challenges for deployment. In contrast, the earliest MoE research introduced a data-dependent, trainable combining method~\cite{jacobs1991adaptive, masoudnia2014mixture}, which aims to decompose complex tasks into simpler sub-tasks, each managed by a dedicated expert. Such works have inspired recent advances in developing modular models called expert specialization~\cite{ma2018modeling, gupta2022sparsely}, providing solutions for deploying large-scale MoE models~\cite{lu2024not} and enabling individual experts to learn and decompose diverse knowledge~\cite{dai2024deepseekmoe}. Inspired by these insights, we propose a specialized fusion method based on MoE integrating multiple audio features to enhance Speech LLMs.

\section{Conclusion}

In conclusion, we propose PaM, a feature fusion method designed to provide a Speech LLM with diverse features from multiple encoders based on users' input prompts. Experimental results indicate that PaM surpasses both single-encoder and multi-encoder baselines across a variety of tasks and datasets. We provide a detailed analysis of the feature importance of different encoders, demonstrating that PaM effectively leverages different encoders and levels of features for distinct tasks. Additionally, we present comprehensive experimental results for the selection and combination of encoders. For future work, we intend to expand the training data and incorporate additional tasks.
\section{Limitations}

Owing to resource constraints, our training data is limited to several hundred hours. It would be preferable to implement our method in larger-scale experiments to facilitate comparison with existing strong Speech LLMs such as Qwen-Audio~\cite{Qwen2-Audio} on a more comprehensive benchmark like Air-Bench~\cite{yang2024air}. Additionally, we train the PaM model from scratch using a predefined list of audio encoders. It would be beneficial to investigate the addition of new encoders to an already trained Speech LLM to enhance its performance on new tasks or in new domains. We leave this for future work.

\section*{Acknowledgements}
This work was supported in part by the National Science Foundation of China (Nos. 62276056 and U24A20334), the Yunnan Fundamental Research Projects (No.202401BC070021), the Yunnan Science and Technology Major Project (No. 202502AD080014), and the Program of Introducing Talents of Discipline to Universities, Plan 111 (No.B16009).

\bibliography{custom}

\appendix
\newpage

\section{Appendix}
\label{sec:appendix}

\subsection{Prompt Example}\label{sec:appendix-prompt-example}

\begin{table}[htp]
    \centering
    \small
    \resizebox{\linewidth}{!}{\begin{tabular}{p{1.5cm}p{4.8cm}}
        \toprule
        \multirow{2}{*}{ASR}
        & 1. Hear the audio clip and transform it into text format. <|AUDIO|>\\
        & 2. Listen to the following audio and create a corresponding text transcript. <|AUDIO|>\\
        \midrule
        Speaker Number & 1. How many speakers' contributions are in this recording? <|AUDIO|>\\
        Verification & 2. What is the number of speakers in this spoken content? <|AUDIO|>\\
        \midrule
        \multirow{2}{*}{AC}
        & 1. Listen to this audio and provide a detailed description. <|AUDIO|>\\
        & 2. Analyze the recording and summarize its contents. <|AUDIO|>\\
        \bottomrule
    \end{tabular}}
    \caption{Examples of prompts for different tasks.}
    \label{table:example}
\end{table}

\subsection{Details of Models and Datasets}\label{sec:appendix-datas-and-model}

In this paper, we leverage multiple audio encoders and LLM to construct the end-to-end speech LLM. Our training dataset is sourced from commonly used open-source datasets, totaling approximately 450 hours of audio data, corresponding to 313,208 samples, as outlined in Table~\ref{tab:whole-training-dataset}. For SNV, we randomly concatenate individual utterances to form new speech signals with the number of speakers ranging from one to four.

\begin{table}[htp]
\small
\centering
\begin{tabular}{l c r c}
\toprule
Data Source & Task & Hours & Sample \\ 
\midrule
Librispeech-clean-100 & \multirow{2}{*}{ASR} & \multirow{2}{*}{100h} & \multirow{2}{*}{28539} \\ 
~\cite{panayotov2015librispeech} & & & \\
\cmidrule(r){1-1}
AMI~\cite{kraaij2005ami} & ASR & 100h & 108502 \\ 
\cmidrule(r){1-1}
Common Voice V4 (Part) & \multirow{2}{*}{SNV} & \multirow{2}{*}{$\sim$150h} & \multirow{2}{*}{137041} \\ 
~\cite{ardila2019common} & & & \\
\cmidrule(r){1-1}
Audio Caption & \multirow{2}{*}{AC} & \multirow{2}{*}{100h} & \multirow{2}{*}{39126} \\ 
~\cite{audiocaps} & & & \\
\bottomrule
\end{tabular}
\caption{The complete training dataset.}
\label{tab:whole-training-dataset}
\end{table}

In our paper, we adopt multiple pre-trained audio encoders and LLMs, and we list the architecture settings for all models we used in our experiments in Table~\ref{tab:model-setting-detail}. Notably, for the Whisper model, we only used its encoder part as an audio feature extractor.

\begin{table*}[htp]
\small
\centering
\begin{tabular}{l r c c c c c l}
\toprule
Audio Encoder Models & Enc Param & Layers & $d_{\text{model}}$ & $d_{\text{ffn}}$ & $d_k$ & $H$ & Norm \\ 
\midrule
\href{https://huggingface.co/openai/whisper-small}{openai/whisper-small} & 88M & 12 & 768 & 3072 & 64 & 12 & Pre \\
\href{https://huggingface.co/microsoft/wavlm-base-plus}{microsoft/wavlm-base-plus} & 94M & 12 & 768 & 3072 & 64 & 12 & Post \\
\href{https://huggingface.co/facebook/wav2vec2-base-960h}{facebook/wav2vec2-base-960h} & 94M & 12 & 768 & 3072 & 64 & 12 & Post \\
\href{https://huggingface.co/openai/whisper-medium}{openai/whisper-medium} & 307M & 24 & 1024 & 4096 & 64 & 16 & Pre \\
\href{https://huggingface.co/microsoft/wavlm-large}{microsoft/wavlm-large} & 315M & 24 & 1024 & 4096 & 64 & 16 & Post \\
\href{https://huggingface.co/facebook/wav2vec2-large-960h}{facebook/wav2vec2-large-960h} & 315M & 24 & 1024 & 4096 & 64 & 16 & Post \\
\midrule
Large Language Models & Lora Param & Layers & $d_{\text{model}}$ & $d_{\text{ffn}}$ & $d_k$ & $H$ & Norm \\
\midrule
\href{https://huggingface.co/Qwen/Qwen2.5-3B}{Qwen/Qwen2.5-3B} & 7M & 36 & 2048 & 11008 & 128 & 16 & Pre \\
\href{https://huggingface.co/Qwen/Qwen2.5-7B}{Qwen/Qwen2.5-7B} & 10M & 28 & 3584 & 18944 & 128 & 28 & Pre \\
\href{https://huggingface.co/meta-llama/Llama-3.2-3B}{meta-llama/Llama-3.2-3B} & 9M & 28 & 3072 & 128256 & 128 & 24 & Pre \\
\href{https://huggingface.co/meta-llama/Llama-3.1-8B}{meta-llama/Llama-3.1-8B} & 13M & 32 & 4096 & 14336 & 128 & 32 & Pre \\
\bottomrule
\end{tabular}
\caption{The settings of the pre-trained model we used in our experiments. For the audio encoder models, we utilize only the encoder component and freeze all parameters. For the LLMs, we freeze the base model parameters and apply LoRA adapters to fine-tune the model.}
\label{tab:model-setting-detail}
\end{table*}

\subsection{Details of Baseline Implementation}\label{sec:appendix-baseline-implement}

In our work, we compare our method against two types of baselines. The first baseline consists of models using a single encoder, while the second baseline involves fusing multiple audio encoders, either as in previous work~\cite{hu2024wavllmrobustadaptivespeech, tang2024salmonn} or through an averaging operation. 

For the single encoder baseline, we train the model using the same settings as in our method. For the second baseline, we train the model using the open-source codebases from \href{https://github.com/microsoft/SpeechT5/tree/main/WavLLM}{WavLLM} and \href{https://github.com/bytedance/SALMONN}{SALMONN}. We integrate the adapter components from these repositories into our code and train the baseline model using our training data, employing the same pre-trained audio encoders and LLMs as in our method. During training, we applied the same hyperparameters as our method. Since we use different encoders and LLMs compared to the baselines, we adjust the dimensions of the adapter to match the specific audio encoder and LLM we used while maintaining other dimensions independent of the audio encoders and LLM unchanged. Notably, we trained SALMONN adapter with query length=32 (as training with the original setting query length=1 failed) to ensure comparable performance with the other baseline methods.

\subsection{Results on Audio Caption with More Metrics}\label{sec:appendix-AC-with-more-metrics}

Due to the multiple evaluation metrics for the AC task,  we evaluated the AC and AC QA tasks using more metrics in Table~\ref{tab:ac-with-more-metrics}, including CIDEr, SPICE (with \href{https://github.com/tylin/coco-caption}{coco-caption toolkit}), FENSE metric, and Sentence-BERT\footnote{We used the FENSE open-source repository GitHub and scored the entire dataset with eval\_system.py. For the SBERT model, we loaded the paraphrase-mpnet-base-v2 model, and for the echecker, we used echecker\_clotho\_audiocaps\_base. However, we encountered a bug when loading the echecker model, which had an unexpected key \textit{encoder.embeddings.position\_ids}. To resolve this, we set strict=False. Additionally, we included results based on similarity using the SBERT model.}. We found that different models exhibited similar performance trends across almost all metrics.

\begin{table*}[!t]
\tiny
\centering
\resizebox{\textwidth}{!}{\begin{tabular}{l c c c c c c c c c c}
\toprule
\multirow{2}{*}{\makecell{Models}} & \multicolumn{4}{c}{AudioCaps} & \multicolumn{4}{c}{AudioCaps QA} \\ 
\cmidrule(r){2-11}
 & METEOR$\uparrow$ & FENSE$\uparrow$ & SBERT$\uparrow$ & CIDEr$\uparrow$ & SPICE$\uparrow$ & METEOR$\uparrow$ & FENSE$\uparrow$ & SBERT$\uparrow$ & CIDEr$\uparrow$ & SPICE$\uparrow$ \\
 \midrule
\rowcolor{black!6} Single-encoder Baselines & & & & & & & & & & \\
\quad- Whisper & 32.96 & 0.108 & 0.596 & 0.431 & 0.158 & 15.04 & 0.105 & 0.402 & 0.205 & 0.083 \\
\quad- WavLM & 27.14 & 0.106 & 0.500 & 0.290 & 0.122 & 12.77 & 0.104 & 0.337 & 0.127 & 0.049 \\
\quad- Wav2Vec2 & 23.81 & 0.096 & 0.414 & 0.205 & 0.095 & 11.20 & 0.092 & 0.282 & 0.073 & 0.038 \\
\midrule
\rowcolor{black!6} Multi-encoder Baselines & & & & & & & & & & \\
\quad- WavLLM & 34.93 & 0.109 & 0.640 & 0.569 & 0.175 & \textbf{16.35} & \textbf{0.108} & \textbf{0.448} & \textbf{0.303} & \textbf{0.109} \\
\quad- SALMONN & 34.86 & 0.109 & 0.631 & 0.542 & 0.158 & 15.97 & 0.108 & 0.432 & 0.254 & 0.093 \\
\quad- Average & 33.22 & 0.108 & 0.615 & 0.471 & 0.166 & 15.53 & 0.108 & 0.425 & 0.229 & 0.093 \\
\midrule
\rowcolor{black!6} PaM & \textbf{35.47} & \textbf{0.111} & \textbf{0.644} & \textbf{0.581} & \textbf{0.183} & 15.70 & 0.108 & 0.428 & 0.267 & 0.087 \\
\bottomrule
\end{tabular}}
\caption{More results based on various metrics on the AC task. The SBERT represents Sentence-BERT.}
\label{tab:ac-with-more-metrics}
\end{table*}

\subsection{PaM with More Encoders}\label{sec:appendix-pam-with-more-encoders}

We added the BEATs encoder to our framework, resulting in a total of four encoders, and found that it significantly improved the performance of our system on the AC task (Table~\ref{tab:result-with-more-encoders}). Although adding the new encoder had limited impact on the AMI and SNV tasks, incorporating the BEATs encoder improved the system's average rank across downstream tasks (Table~\ref{tab:rank-with-more-encoders}). We plan to conduct additional experiments with the EAT encoder~\cite{ijcai2024p421} in future work.

\begin{table*}[htp]
\tiny
\centering
\resizebox{\textwidth}{!}{\begin{tabular}{l c c c c c c}
\toprule
\multirow{2}{*}{Prompt-aware Mixture (PaM)} & \multicolumn{2}{c}{LibriSpeech} & AMI & SNV & AudioCaps & AudioCaps QA \\ 
\cmidrule(r){2-7}
 & WER(clean)$\downarrow$ & WER(other)$\downarrow$ & WER$\downarrow$ & Acc$\uparrow$ & METEOR$\uparrow$ & METEOR$\uparrow$ \\ 
\midrule
\quad- PaM & 3.65 & 7.07 & 12.79 & 42.8\% & 35.47 & 15.70 \\
\quad- PaM (BEATs) & 3.76 & 7.22 & 13.11 & 49.2\% & 35.70 & 16.36 \\
\bottomrule
\end{tabular}}
\caption{Results of PaM with more encoders.}
\label{tab:result-with-more-encoders}
\end{table*}

\begin{table*}[htp]
\centering
\resizebox{\textwidth}{!}{\begin{tabular}{l c c c c c c c c c}
\toprule
Model & Whisper & WavLM & Wav2Vec2 & WavLLM & SALMONN & Average & PaM (audio-based) & PaM (prompt-aware) & PaM (BEATs) \\
\midrule
AVG Rank & 7.7 & 7.5 & 7.3 & 4.5 & 5.0 & 5.3 & 3.5 & 2.5 & 1.7 \\
\bottomrule
\end{tabular}}
\caption{Average result rank in all downstream tasks of different models.}
\label{tab:rank-with-more-encoders}
\end{table*}

\subsection{Details of Training and Inference Parameters}\label{sec:appendix-training-inference-param}

We train our model for five epochs with a learning rate of 5e-5, 2000 warmup steps, and bf16 precision. We freeze all encoders and the LLM, only train adapters and the fusion modules. For the LLM, we apply LoRA~\cite{hu2022lora} with a rank of 32 and an alpha of 64, adding LoRA only to the q\_proj and k\_proj. Each task has the same probability during training. During the inference stage, we select the last checkpoint on the validation set and perform greedy search.

\subsection{WavLLM and PaM with Larger LLM}\label{sec:appendix-wavllm-and-pam-larger-llm}

We experiment with WavLLM (concatenation) and PaM under a larger scale LLM based on Qwen2.5-7B and Llama3.1-8B, as shown in Table~\ref{tab:Wavllm-vs-PaM-scaling}. We found that PaM consistently outperforms concatenation on LibriSpeech. However, in noisier ASR scenarios such as AMI, concatenation performs better. On tasks like SNV, AC, and AC QA, concatenation’s performance is not stable. In the Qwen-based speech large language model, the performance on these three tasks is better than PaM, but in the Llama-based model, the performance on these tasks is significantly worse than PaM.

\begin{table*}[htp]
\tiny
\centering
\resizebox{\textwidth}{!}{\begin{tabular}{l c c c c c c}
\toprule
\multirow{2}{*}{\makecell{Models}} & \multicolumn{2}{c}{LibriSpeech} & AMI & SNV & AudioCaps & AudioCaps QA \\ 
\cmidrule(r){2-7}
& WER(clean)$\downarrow$ & WER(other)$\downarrow$ & WER$\downarrow$ & Acc$\uparrow$ & METEOR$\uparrow$ & METEOR$\uparrow$ \\ 
\midrule
\rowcolor{black!6} Qwen2.5-7B & & & & & & \\
\quad- WavLLM & 5.13 & 10.81 & 13.66 & 51.1\% & 36.27 & 16.55 \\
\quad- PaM & 3.68 & \phantom{0}8.36 & 15.26 & 43.7\% & 35.46 & 15.71 \\
\midrule
\rowcolor{black!6} LLaMA3.1-8B & & & & & & \\
\quad- WavLLM & 6.38 & 14.08 & 13.95 & \phantom{0}3.1\% & 34.91 & 14.76 \\
\quad- PaM & 4.85 & \phantom{0}8.87 & 15.01 & 50.8\% & 35.81 & 15.82 \\
\bottomrule
\end{tabular}}
\caption{Results based on our PaM adapter and WavLLM with larger LLMs.}
\label{tab:Wavllm-vs-PaM-scaling}
\end{table*}

We note that the concatenation method in the Llama-based model performs significantly worse on SNV, with only 3.1\% accuracy. This is because it becomes difficult to follow the instructions of the SNV task during the inference stage. After further experiments, we found that the concatenation method becomes progressively less effective on SNV. This suggests that the method struggles to achieve a balance between multitasking as training progresses.

\subsection{Ablation Experiments of Routing Method}\label{sec:appendix-ablation-routing}

We adopt a prompt-aware routing method to better utilize the information in the prompt based on the LLM, as described in Equations~\ref{eq:7} and~\ref{eq:8}. To further evaluate the impact of different routing strategies, we conducted ablation experiments on various forms of routing methods, as presented in Table~\ref{tab:more-results-based-on-routing-strategy}. 

\begin{itemize}
\item \textbf{Audio-based Routing Method.} The routing method employed in most of the MoE models, such as Switch Transformers~\cite{fedus2022switch}, DeepSeekMoE~\cite{dai2024deepseekmoe}, and MoWE~\cite{zhang2024moweaudiomultitaskaudiollmsmixture}, which use the current layer input as the routing module input, and optimize the routing module directly based on the loss of outputs. However, this approach ignores information from task labels. 
\begin{align}
 \label{eq:Moe-routing}
 \text{G} &= \text{Top-}k(\text{Softmax}(\mathbf{X}))
\end{align}

We set $k=1$ in the $\text{Top-}k$ function, consistent with the configuration used in our PaM setup. In our model, the MoE layer is positioned after multiple encoders, since we use the fused features from multiple encoders as the routing inputs.
\begin{align}
 \label{eq:Moe-routing-input}
 \mathbf{X} &= \text{FFN} (\text{Concat}(\mathcal{H}^\text{last}))
\end{align}
\end{itemize}

In addition, we also performed ablation experiments with our PaM routing method.

\begin{itemize}
\item \textbf{Without Shared Expert.} We maintain the full model configuration but remove the shared expert.
\item \textbf{Without Task Label.} We retain the use of task-related information extracted from the LLM prompt as input to the routing module, without any additional labeling information. Specifically, we remove the auxiliary loss term $\mathcal{L}_\text{G}(\text{P}(\text{Task}|\mathbf{H}_\text{prompt}), \text{Task})$ in Equation~\ref{eq:10}. In contrast to the audio-based routing method, this variant of the PaM routing method uses the prompt feature $\mathbf{X}_\text{prompt}$ as input but without the task label.
\end{itemize}

\begin{table*}[t]
\centering
\resizebox{\textwidth}{!}{\begin{tabular}{l l l l c c c c}
\toprule
\multirow{2}{*}{Models} & \multicolumn{2}{c}{LibriSpeech} & AMI & SNV & AudioCaps & AudioCaps QA & \multirow{2}{*}{AVG Rank$\downarrow$} \\ 
\cmidrule(r){2-7}
 & WER(clean)$\downarrow$ & WER(other)$\downarrow$ & WER$\downarrow$ & Acc$\uparrow$ & METEOR$\uparrow$ & METEOR$\uparrow$ & \\ 
\midrule
MoE (audio-based) & 4.12 & \phantom{0}8.32 & 11.21 & 26.0\% & 35.66 & 15.61 & 3.17\\
PaM (with one expert) & 4.27 & 10.04 & 13.77 & 45.8\% & 35.76 & 16.47 & 2.83 \\
PaM (without shared) & 4.31 & \phantom{0}7.89 & 13.43 & 45.2\% & 35.27 & 16.35 & 3.33 \\
PaM (without task label) & 3.97 & \phantom{0}7.97 & 13.14 & 43.5\% & 35.54 & 15.42 & 3.17\\
PaM (ours) & 3.65 & \phantom{0}7.07 & 12.79 & 42.8\% & 35.47 & 15.70 & 2.50 \\
\bottomrule
\end{tabular}}
\caption{Ablation results on our routing method and the results based on the audio-based routing method}
\label{tab:more-results-based-on-routing-strategy}
\end{table*}

We found that the PaM routing method outperforms audio-based routing on most ASR and AC QA tasks, especially SNV tasks. This suggests that, in our setting, PaM is superior to the basic MoE routing method for fusing multi-encoder features. 

For the PaM without shared experts, we found that it still outperforms the single model baseline and maintains better or comparable performance compared to the multi-encoder baseline on almost all tasks. Compared to PaM without shared experts, PaM with shared experts gains on several tasks but is slightly weaker on AC QA and SNV. This suggests that while shared experts may slightly degrade performance on a few tasks, they can significantly improve the overall effectiveness of the PaM model. 

Compared to PaM without task labels and PaM, we found that PaM achieved improvement on most of the tasks, which further illustrates the effectiveness of incorporating task information in the prompt when handling multiple downstream tasks.

\subsection{Initialization setup for LLM}\label{sec:appendix-init-setup-for-LLM}

We initialize the LLM module in PaM from the open-source base model, following the setup used widely in prior work~\cite{hu2024wavllmrobustadaptivespeech, ma2024embarrassingly, tang2024salmonn, Rubenstein2023AudioPaLMAL}. An interesting alternative is to initialize our LLM module with the LLM module in a pretrained multimodal model like Qwen-Audio. We are not adapting such a setup because Qwen-Audio’s feature alignment was specifically designed for Whisper’s encoder, which can potentially "overfit" to features from Whisper. Our new experiments in Table~\ref{tab:cosine-sim-encoders} reveal that the feature spaces of Wav2Vec2 and WavLM are relatively similar, while Whisper's feature space shows greater divergence. This pattern is also reflected in the weight distribution in Figure~\ref{fig:hidden-layer-importance}, where Wav2Vec2 and WavLM appear more closely aligned and significantly different from Whisper. Therefore, using other encoders and restructuring the projector module would still require re-adapting the LLM to comprehend new features.

\begin{table*}[htp]
\centering
\resizebox{\textwidth}{!}{\begin{tabular}{l c c c c c c c c c c c c c}
\toprule
Cosine similarity & emb & 1 & 2 & 3 & 4 & 5 & 6 & 7 & 8 & 9 & 10 & 11 & 12 \\
\midrule
Whisper \& WavLM & -0.91 & -0.76 & -0.79 & -0.83 & -0.83 & -0.85 & -0.89 & -0.89 & -0.89 & -0.84 & -0.69 & -0.38 & -0.85 \\
Whisper \& Wav2Vec2 & -0.90 & -0.73 & -0.77 & -0.80 & -0.81 & -0.84 & -0.89 & -0.89 & -0.89 & -0.85 & -0.75 & -0.69 & -0.86 \\
WavLM \& Wav2Vec2 & \ 0.99 & \ 0.13 & \ 0.26 & \ 0.44 & \ 0.52 & \ 0.60 & \ 0.95 & \ 0.99 & \ 0.99 & \ 0.52 & \ 0.06 & -0.38 & \ 0.48 \\
\bottomrule
\end{tabular}}
\caption{The cosine similarity of the hidden states between the layers of different encoders.}
\label{tab:cosine-sim-encoders}
\end{table*}

\subsection{All Detailed Results}

The detailed results of our experiments with multiple encoders are summarized in Table~\ref{tab:results-multiple-encoder}. We observe that, in most cases, the audio encoder that performed well on a single task also enhanced the performance of the fusion model on that task. In cases where performance degradation occurs on a specific task when using the corresponding encoder, the fusion model consistently includes the HuBERT audio encoder, suggesting that incorporating the HuBERT model may have a detrimental effect. This could be attributed to the fact that the HuBERT model is trained on a smaller pre-trained dataset compared to other audio encoders. Notably, even in this case, fusing four audio encoders yields comparable results to fusing three encoders on the AVG Rank, indicating that incorporating more encoders can still lead to performance improvements.

\begin{table*}[htp]
\centering
\resizebox{\textwidth}{!}{\begin{tabular}{c c c c c c c c c c c c c c}
\toprule
\multirow{2}{*}{Encoders} & \multirow{2}{*}{Whisper} & \multirow{2}{*}{WavLM} & \multirow{2}{*}{Wav2Vec2} & \multirow{2}{*}{HuBERT} & \multicolumn{2}{c}{LibriSpeech} & AMI & SNV & AudioCaps & AudioCaps QA & \multicolumn{2}{c}{Avg} & \multirow{2}{*}{AVG Rank} \\ 
\cmidrule(r){6-13}
& & & & & WER(clean)$\downarrow$ & WER(other)$\downarrow$ & WER$\downarrow$ & Acc$\uparrow$ & METEOR$\uparrow$ & METEOR$\uparrow$ & $\downarrow$ & $\uparrow$ & \\ 
\midrule
1 & $\surd$ & - & - & - & 9.61 & 16.73 & 16.27 & 18.8\% & 32.96 & 15.04 & 14.20 & 22.27 & 11.67 \\
1 & - & $\surd$ & - & - & 5.59 & 10.57 & 18.97 & 41.4\% & 27.14 & 12.77 & 11.71 & 27.10 & 11.50 \\
1 & - & - & $\surd$ & - & 4.30 & \phantom{0}9.46 & 26.69 & 39.0\% & 23.81 & 11.20 & 13.48 & 24.67 & 11.67 \\
1 & - & - & - & $\surd$ & 7.47 & 13.85 & Fail & 31.1\% & 23.94 & 11.28 & Fail & 22.10 & 14.00 \\
\midrule
\rowcolor{black!10} \multicolumn{5}{l}{Best-1} & 4.30 & \phantom{0}9.46 & 16.27 & 41.4\% & 32.96 & 15.04 & 13.13 & 24.04 & \\
\midrule
2 & $\surd$ & $\surd$ & - & - & 5.07 & \phantom{0}9.50 & 13.59 & 49.5\% & \textbf{35.47} & 16.16 & \phantom{0}9.38 & 33.71 & \phantom{0}6.00 \\
2 & $\surd$ & - & $\surd$ & - & 3.82 & \phantom{0}7.55 & 13.51 & 38.4\% & 35.43 & 15.55 & \phantom{0}8.29 & 29.79 & \phantom{0}5.83  \\
2 & $\surd$ & - & - & $\surd$ & 4.37 & \phantom{0}9.70 & 14.79 & 43.2\% & 35.33 & 16.62 & \phantom{0}9.62 & 31.72 & \phantom{0}6.50 \\
2 & - & $\surd$ & $\surd$ & - & 3.17 & \phantom{0}6.76 & 19.60 & \textbf{61.2\%} & 31.70 & 14.89 & \phantom{0}9.84 & 35.93 & \phantom{0}5.83 \\
2 & - & $\surd$ & - & $\surd$ & 3.96 & \phantom{0}7.64 & 22.24 & 31.7\% & 30.28 & 13.99 & 11.28 & 25.32 & 10.33 \\
2 & - & - & $\surd$ & $\surd$ & 3.27 & \phantom{0}7.29 & 21.43 & 35.8\% & 29.85 & 14.62 & 10.66 & 26.76 & \phantom{0}9.00 \\
\midrule
\rowcolor{black!10} \multicolumn{5}{l}{Best-2} & 3.17 & \phantom{0}6.76 & 13.51 & 61.2\% & 35.47 & 16.62 & \phantom{0}9.85 & 36.65 & \\
\midrule
3 & $\surd$ & $\surd$ & $\surd$ & - & 3.65 & \phantom{0}7.07 & \textbf{12.79} & 42.8\% & \textbf{35.47} & 15.70 & \phantom{0}\textbf{7.83} & 31.32 & \phantom{0}4.00 \\
3 & $\surd$ & $\surd$ & - & $\surd$ & 4.42 & 10.26 & 13.47 & 47.4\% & 35.32 & 16.62 & \phantom{0}9.38 & 33.11 & \phantom{0}6.50 \\
3 & $\surd$ & - & $\surd$ & $\surd$ & 4.58 & \phantom{0}6.99 & 15.11 & 59.1\% & 35.33 & 16.62 & \phantom{0}8.89 & \textbf{37.02} & \phantom{0}5.50 \\
3 & - & $\surd$ & $\surd$ & $\surd$ & \textbf{3.09} & \phantom{0}\textbf{6.64} & 22.80 & 26.0\% & 31.90 & 15.14 & 10.84 & 24.35 & \phantom{0}7.67 \\
\midrule
\rowcolor{black!10} \multicolumn{5}{l}{Best-3} & 3.09 & \phantom{0}6.64 & 12.79 & 59.1\% & 35.47 & 16.62 & \phantom{0}9.24 & 31.45 & \\
\midrule
4 & $\surd$ & $\surd$ & $\surd$ & $\surd$ & 3.94 & \phantom{0}7.28 & 14.06 & 57.3\% & 35.37 & \textbf{16.79} & \phantom{0}8.43 & 36.49 & \phantom{0}4.00 \\
\bottomrule
\end{tabular}}
\caption{Detailed results of incorporating different combinations of audio encoders.}
\label{tab:results-multiple-encoder}
\end{table*}

\subsection{Prompts in Training and Inference Stage}\label{sec:appendix-prompts-in-training-inference}

In practice, large speech-language models typically address downstream tasks using prompts that are semantically explicit but textually diverse, as illustrated in Figure~\ref{fig:layer_importance}(a). Unlike classical MoE routing methods that rely on hidden states, or other approaches such as predefined task labels (e.g., “Task ID = 3”)~\cite{kudugunta2021beyond}, and random routing~\cite{zuo2021taming}, our model leverages natural language prompts to route inputs to the appropriate expert.

These prompts vary in phrasing but convey the same task intent. We utilize the semantic understanding capabilities of a pretrained LLM to extract task information from these prompts, rather than depending on fixed task labels. As an example, in our ASR task, we trained the routing module using 50 diverse prompts (Figure~\ref{fig:layer_importance}(b)). During evaluation, the model was tested on 200 prompts, 150 of which were unseen during training (Figure~\ref{fig:layer_importance}(c)). Our prompt-based router, powered by the LLM, achieved 100\% expert selection accuracy, demonstrating strong generalization to previously unseen prompts.

\begin{figure*}[htp]
  \centering
  \centerline{\includegraphics[scale=1.0]{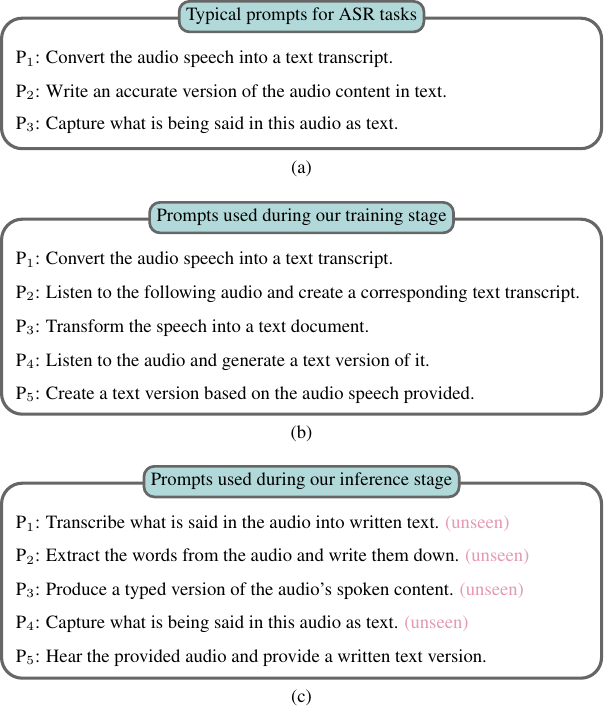}}
  \caption{Example of prompts.}
\label{fig:layer_importance}
\end{figure*}

\subsection{Efficiency}\label{sec:appendix-efficient}

The low decoding efficiency of LLMs is due to repeated invocations of the entire decoder stack during autoregressive generation. Several methods have been proposed to address this issue, including speculative sampling~\cite{leviathan2023fast} and KV-cache compression~\cite{zhang2023h2o}. Recently, researchers have explored using LLMs as encoders~\cite{luo2025beyond}, thereby leveraging their knowledge while improving decoding efficiency. Additionally, enhancing the computational efficiency and performance of adapters in end-to-end speech models through optimized FFN dimension design~\cite{zheng2023partialformer} represents a viable solution.

\end{document}